\documentclass[12pt,preprint]{aastex} 

\tighten 

\def\lsim{\lower.5ex\hbox{$\; \buildrel < \over \sim \;$}} 
\def\gsim{\lower.5ex\hbox{$\; \buildrel > \over \sim \;$}} 
\def\lax {\ifmmode{_<\atop^{\sim}}\else{${_<\atop^{\sim}}$}\fi} 
\def\gax {\ifmmode{_>\atop^{\sim}}\else{${_>\atop^{\sim}}$}\fi} 
 
\def\gtorder{\mathrel{\raise.3ex\hbox{$>$}\mkern-14mu 
\lower0.6ex\hbox{$\sim$}}} 
\def\ltorder{\mathrel{\raise.3ex\hbox{$<$}\mkern-14mu 
\lower0.6ex\hbox{$\sim$}}}

\def\pmb#1{\setbox0=\hbox{#1}%
\kern-0.015em\copy0\kern-\wd0 
\kern0.03em\copy0\kern-\wd0 
\kern-0.015em\raise0.0433em\box0 }

\begin{document} 

\title{Rayleigh - Taylor Gravity Waves and  Quasiperiodic Oscillation Phenomenon 
in X-ray Binaries} 

\author{Lev Titarchuk \altaffilmark{1,2}}

\altaffiltext{1}{George Mason University/CEOSR/NRL;
lev@xip.nrl.navy.mil}
\altaffiltext{2}{NASA Goddard Space Flight Center, code 661, 
Laboratory for High Energy
Astrophysics, Greenbelt MD 20771; lev@lheapop.gsfc.nasa.gov}

\shorttitle{Rayleigh- Taylor Gravity Waves}
\shortauthors{TITARCHUK}

\begin{abstract} 
Accretion  onto compact objects in X-ray binaries (black hole, neutron star (NS), white dwarf)
is characterized  by  non-uniform flow density profiles.
Such an effect of heterogeneity in presence of gravitational forces
and pressure gradients     exhibits 
Raylegh-Taylor gravity waves (RTGW). 
They should be seen as quasioperiodic wave oscillations (QPO) of the
accretion  flow in the transition (boundary) layer between the Keplerian disk
and the central object. 
In this paper I show that the main QPO frequency, which is very close to the Keplerian frequency,
is split into separate frequencies (hybrid and low branch) under the influence of the 
gravitational forces in the rotational frame of reference. The RTGWs must be present and the related QPOs should be
detected in any  system where the gravity, buoynancy and Coriolis force effects
 cannot be excluded (even in the Earth and solar environments). 
The observed low  and high QPO frequencies are an intrinsic signature of the RTGW.
I elaborate the conditions for the density profile  when the RTGW oscillations are stable. 
A comparison of the inferred QPO frequencies with QPO observations is presented. 
I find that  hectohertz frequencies detected
from NS binaries can be identified as  the RTGW low branch frequencies. 
I also predict that an observer can see the double NS spin frequency 
during the NS  long (super) burst events when the pressure gradients and buoyant forces
are suppressed.
The Coriolis force is the only force
which acts in the rotational frame of reference and its presence 
 causes perfect coherent pulsations
with a frequency twice of  the NS spin.
The QPO observations of neutron binaries have established that 
the high QPO frequencies do not go beyond of 
the certain upper limit.  
I  explain this observational effect as a result of 
the density profile inversions.
Also I demonstrate that a  particular problem of the gravity waves in the rotational frame of reference
in the approximation of very small pressure gradients is reduced to the problem of the classical oscillator 
in the rotational frame of reference  which was previously introduced   and applied for the interpretation 
of kHZ  QPO observation by Osherovich \& Titarchuk. 

\end{abstract} 


\section{Introduction} 
The theory of oscillations of  rotating fluids of variable density 
[Rayleigh- Taylor (R-T) effect] was developed in detail by Chandrasekhar (1961), hereafter C61. 
A large variety of the magnitohydrodynamic (MHD) problems,
 including the stability of inviscid and viscous fluids in the case of two
uniform  layers separated by a horizontal boundary with and without rotation as well as the magnetic
field effects were analyzed using perturbation technique. 
The simplest case of a one-dimensional gravitational force was studied.  
A similar analysis was also  implemented for the case of an exponentially varying density.
It follows from C61 that  the quasiperiodic oscillations (QPO) with the twin ``kilohertz''  
frequencies represent the main frequencies of the stable gravity waves in the rotational frame of reference.
It also as  follows from C61  that the twin ``kilohertz''  frequencies should be on the order of the Keplerian frequency.
  
In this {\it Paper} I  present the results of  the  study of the R-T effect  for  a particular case of 
 fluid oscillations  in an accretion flow under influence of  a central gravitational force.       
This  particular R-T analysis is important in  view of  the high and low frequency detection by        
the Rossi X-ray Timing Explorer (RXTE) in a number of low mass X-ray
binaries (Strohmayer et al.  1996, van der Klis et al. 1996), black hole candidate sources (Morgan, Remillard \& Greiner 
1997; Strohmayer 2001a,b; Remillard 2002)
and by Extreme Ultraviolet Explorer, Chandra X-ray observatory and optical observations
in white dwarfs (Mauche 2002 and Woudt \& Warner 2002).
The presence of two observed peaks with frequencies $\nu_1$ and $\nu_2$ in
the upper part of the power spectrum became a natural starting point
in modeling the phenomena.   In NS binaries, for example Sco X-1,   the lower
frequency part of the power spectrum, contains two horizontal branch
oscillation (HBO) frequencies $\nu_{HBO}\sim 45$ Hz and
$\nu_{2HBO}\sim 90$ Hz (probably the second harmonic of $\nu_{HBO}$)
which slowly increase with the increase of $\nu_1$ and $\nu_2$ (van der
Klis et al. 2000). 
{\it Any plausible model faces the challenging task
of describing the dependences of the peak separation $\Delta\nu=\nu_2-\nu_1$ on $\nu_1$ and $\nu_2$} .
Attempts have been made to relate $\nu_1$
and $\nu_2$ and the peak separation $\Delta \nu= \nu_2-\nu_1$ with the
neutron star (NS) spin. 
In  the sonic point beat frequency  model  by Miller, Lamb \&
Psaltis (1998) the kHz peak separation $\Delta \nu$ is considered to be
close to the NS spin frequency and thus $\Delta \nu$ is predicted to
be constant. 
However observations
of kHz QPOs in a number of binaries (Sco X-1, 4U 1728-34, 4U 1608-52,
4U 1702-429 and etc) show that the peak separation decreases
systematically when the high [kilohertz (kHz)] frequencies increase (for  a recent review see van der
Klis 2000, hereafter VDK).  For Sco X-1 VDK found that the
peak separation of kHz QPO frequencies changes from 320 Hz to 220 Hz
when the lower kHz peak $\nu_1$ changes from 500 Hz to 850 Hz.

The correlation between high frequency (lower kHz frequency) and low frequency 
(broad noise component) QPOs previously found by Psaltis, Belloni \& van der Klis (1999) for black hole (BH) and 
 neutron star (NS) systems has been recently extended 
 over two orders of magnitude by Mauche (2002) to white dwarf (WD) binaries.
{\it Accepting the reasonable assumption that the same mechanism produces the QPO in  
WD, NS and BH binaries, one can argue that the data exclude relativistic models and, beat frequency models 
as well as any model requiring either the presence or absence of a stellar surface or a strong magnetic field.}  

The transition layer model (TLM) was introduced by Titarchuk, Lapidus
\& Muslimov (1998), hereafter TLM98, to explain the dynamical
adjustment of a Keplerian disk to the innermost sub-Keplerian boundary
conditions (it is  at the star surface for NSs and WDs). 
 TLM98 argued that a shock should
occur where the Keplerian disk adjusts to the sub-Keplerian flow. 
Thus the transition layer bounded between the  sub-Keplerian boundary and the adjustment radius can undergo
various type of oscillations under the influence of the gas, radiation,  magnetic pressure and gravitational 
force.
Osherovich \& Titarchuk (1999), hereafter OT99, suggested that the phenomological model of a
one-dimensional classical oscillator in the rotational frame of reference could explain the observed
correlations between twin kHz frequencies $\nu_1$, $\nu_2$ and the HBO frequencies.
They further suggested that the  oscillations of the fluid element that bounced from the
disk shock region (at the adjustment radius) would be seen
as two independent oscillations parallel and perpendicular to the disk  plane respectively.
This is due to the presence of a Coriolis force in the
magnetospheric rotational frame of reference. 
In this paper I show that when the pressure gradients can be neglected,
 the problem of the Rayleigh-Taylor wave oscillations (buoynancy effect) 
in the rotational frame of reference   is reduced to the OT99 formulation (see \S 3).
This result provides a solid basis for application of the OT99 model for interpretation of the QPO 
phenomena observed in X-ray binaries. 

The main goal of this {\it Paper} is to demonstrate that there is an inevitable effect of the gravity wave
oscillations in the heterogeneous fluid of the accretion flow near compact objects.
In \S 2, I formulate  the problem of the gravity wave propagation in the bounded medium in the rotational frame of
reference.   In \S 3 I  present analysis and  solutions of this problem using a perturbation method 
in the context of three dimensional periodic waves with assigned wave
numbers. Applications of the gravity modes and their relations 
with the QPO observations
is presented in \S 4. Summary and conclusions are drawn in \S 5. 

\section{The  Problem of Gravity Wave Propagation in a Bounded Medium}  

Lord Rayleigh (1883) has treated the non-rotating inviscid case of the present problem. He developed a general theory for any
density configuration $\rho_0(z)$ (the z axis being the upward drawn vertical).  Rayleigh's treatment of inviscid superposed
fluids was extended by Bjerkness et al. (1933) to include the influence of rotation.
With an assumption of the gravitational force directed vertically, Hide  (1956) developed the theory 
for any density  $\rho_0(z)$ and viscosity $\mu_0(z)$ profiles and any $\delta$
(where $\delta$ is an angle between the rotational axis and the vertical).
In this study we consider a case of an inviscid and incompressible fluid.

The equation of relative motion appropriate to the problem is
\begin{equation}
\rho\frac{\partial u_i}{\partial t}+ \rho u_i \frac{\partial}{\partial x_j}u_i-2\rho\Omega\epsilon_{ijk}u_js_k=
-\frac{\partial p}{\partial x_j} -g\rho e_i,
\end{equation}
where the fluid is supposed to rotate uniformly about an axis whose direction is specified by the unit 
vector ${\bf s}={\bf \Omega}/\Omega$.  The tensor notation follows the summation convention  and the unit vector ${\bf e}$ 
in the radial direction is introduced.  
In equation (1) $\rho$ denotes the density,  $\Omega$ the angular velocity of rotation,
$u_i$ the $i$th component of the (Eulerian) velocity vector, p is the pressure and $g$ is the acceleration due to gravity.
This  term $g$ can represent the effective gravity which includes the centrifugal and radiation  pressure forces.
 For an incompressible fluid the continuity
equation is
\begin{equation}
\frac{\partial u_i}{\partial x_i}=0.
\end{equation}
Because (similar to C61)diffusion effects  are ignored in this analysis, an individual fluid element
  retains the same
density throughout its motion. Hence it is required that
\begin{equation}
\frac{D\rho}{Dt}=\frac{\partial\rho}{\partial t}+u_j\frac{\partial \rho}{\partial x_j}=0.
\end{equation}
Because the equilibrium situation in the comoving frame  is a static one it is characterized by $u_i=0$.
We now assume the equilibrium situation to be slightly disturbed, so that $u_i\neq0.$ So we shall write
\begin{equation}
\rho=\rho_0(r,z)+\delta\rho(x,y,z), ~~~ p=p_0(z)+\delta p(x,y,z,t).
\end{equation}
and treat $u_i$, $\delta\rho$, and $\delta p$ as quantities of the first order of smallness 
so that products of such quantities can be ignored. 

We are interested in the study of the gravity wave oscillations in the disk transition layer 
where the accretion flow is sub-Keplerian and the temperature of the flow is of order 5 keV or more 
(see cartoon diagram of the system  in Titarchuk, Osherovich 
\& Kuznetsov 1999, hereafter TOK,  Fig. 1).  These temperatures of order of 5 keV are a representative values for 
 plasma temperatures inferred from X-ray spectra during  kHz QPO events
(see more in Titarchuk, Bradshaw \& Wood 2001 and Titarchuk \& Wood 2002). 
   
At this stage we introduce the Cartesian coordinate system. 
We take the $z$ axis to be in the direction of the (upward) vertical and the $x$  axis to be such that the $(x,~z)$ plane contains the angular
velocity vector ${\bf \Omega}$ which then has components $(\Omega_x,0,\Omega_z)$. 
In the general case the gravitational force has  components $g(x/R, y/R, z/R)$ where $R=(x^2+y^2+z^2)^{1/2}$. 

If $(u,v, w)$ are the components of the velocity ${\bf u}$, then on substituting in equations (1), (2) and (3) we find
\begin{eqnarray}
\rho_0\frac{\partial u}{\partial t}-2\rho_0 \Omega_zv= -\frac{\delta p}{\delta x} -g_x\delta \rho,\\
\rho_0\frac{\partial v}{\partial t}-2\rho_0 \Omega_x w +2\rho_0 \Omega_z u= -\frac{\delta p}{\delta y} -g_y\delta \rho,\\
\rho_0\frac{\partial w}{\partial t}+2\rho_0 \Omega_xv= -\frac{\delta p}{\delta z} -g_z\delta \rho,\\ 
\frac{\partial u}{\partial x} + \frac{\partial v}{\partial y}+ \frac{\partial w}{\partial z}=0,\\ 
\frac{\partial \delta\rho}{\partial t} + u\frac{\partial \rho_0}{\partial x}+ 
v\frac{\partial\rho_0 }{\partial y} +w\frac{\partial \rho_0}{\partial z}=0.
\end{eqnarray}
One can see that the variation of $\rho$ is ignored in all terms except for the ones representing the buoyancy force
(see C61 for details).
\section{Analysis of various types of the solutions of the problem}

In order to illustrate the buoyancy (R-T) effect we consider the simplest solutions of the problem
including those which already exist in the literature.
\par
\noindent
{\it Case A}
\par
In the case when we allow
the vector ${\bf \Omega}$ to rotate around the $z$ axis at angle $\delta$,
we  introduce the cylindrical coordinate system, namely  
taking the $z$ axis in the direction of the (upward) vertical, the $x$  axis as the radial axis  and the $y$ axis as
the azimuthal axis in the horizontal plane.  Then the vector ${\bf \Omega}$  has components $(\Omega \sin \delta,0,\Omega\cos\delta)$
and gravitational force vector has components  $g(r/R, 0, z/R)$ where  $r=(x^2+y^2)^{1/2}$.

(I)~With  assumptions that $\rho_0$ is only a function of $R$ 
 the scale height of $\rho_0$ is order of $R$ we obtain that 
$\delta \rho=[(d\rho/dx)dx+(d\rho/dz)dz)\sim\rho_0/R[(r/R)\chi+(z/R)\zeta]$ where $\chi=dx$, $\zeta=dz$
are the radial and vertical components of displacement respectively and  consequently
the radial  and vertical components of the perturbation 
velocity  are $u= d\chi/dt$, $w=d\zeta/dt$. 
Thus the  vector ${\bf\cal G}=\delta \rho(g_x, 0, g_z)$ is transformed into 
${\bf\cal G}=\rho_0(g/R)[(r/R)(r\chi/R+z\zeta/R),~0,~(z/R)(r\chi/R+z\zeta/R]$.

Furthermore if we neglect the effects of pressure gradients,   the set of equations (5-7) can be reduced 
to the system where the right hand side consists of only the vector ${\bf\cal G}$. 
Then, for radial, azimuthal and vertical displacements $\chi$, $\Upsilon$, $\zeta$ this system takes the form:
 \begin{eqnarray}
\ddot\chi-2\Omega \cos\delta\dot\Upsilon= -(g/R)(r/R)(r\chi/R+z\zeta/R),\\
\ddot\Upsilon+2\Omega\cos\delta \dot\chi-2 \Omega\sin\delta \dot\zeta = 0,~~~~~~~~~~~~~~~~~~~~~\\
~~~~~~~~~~~~~~~~~~~~~~~~~\ddot \zeta+2 \Omega\sin\delta\dot\Upsilon= -(g/R)(z/R)(r\chi/R+z\zeta/R).
\end{eqnarray}
 OT99 have already analyzed the solution of Eqs. (10-12)
(see Eqs. 2-4,  in OT99) in the case of $z/R\ll0$  and $\delta\ll1$. 
 They found that in the rotational frame of reference, 
the radial oscillation with the main frequency
 $\omega_{\rm K}=(g/R)^{1/2}$  is split to the oscillations taking place near 
 the horizontal (disk) plane ($\chi\Upsilon$) with 
the hybrid frequency $\omega_h=(\omega_{\rm K}^2+4\Omega^2)^{1/2}$ and  to oscillations taking place near
 the vertical plane 
($\Upsilon\zeta$) with the low branch frequency $\omega_L=2\Omega(\omega_{\rm K}/\omega_h)\sin\delta$.

 For  $z\sim r$
 the dispersion equation for the frequency $\omega$
\begin{equation}
\omega^2[\omega^4-2(\omega_\ast^2+2\omega^2)\omega^2+\omega_\ast^4+4\Omega^2\omega_\ast^2
+2\Omega^2\omega_\ast^2\sin2\delta]=0
\end{equation}
besides the nonoscillatory mode ($\omega=0$), describes two oscillatory  eigenmodes.
For $\delta\ll1$ they are $\omega_1=\omega_\ast=(g/2R)^{1/2}=\omega_{\rm K}/\sqrt2$ 
and $\omega_2=(\omega_{\ast}^2+4\Omega^2)^{1/2}=(\omega_{\rm K}^2/2+4\Omega^2)^{1/2}$. 

The relation of the model eigenfrequencies $\nu_L=\omega_L/2\pi$, $\nu_{\rm K}=\omega_{\rm K}/2\pi$ and 
$\nu_h=\omega_h/2\pi$ with the QPO observed frequencies: horizontal branch oscillation freqiencies 
$\nu_{HBO}$, kHz frequencies were studied by OT99; TOK, 
Kuznetsov \& Titarchuk (2002); and Titarchuk (2002), hereafter T02. It is worth noting that the relation between
$\nu_{\rm K}$, $\Omega$, $\nu_h$ and $\nu_L$ predicts the existence of the invariant quantity $\delta$.
(e.g. Titarchuk \& Osherovich 2001).
T02 calculated $\delta$ and its uncertainty of $\delta$  finding that the
inferred $\delta-$values are consistent with being constant at least for four Z sources, Sco X-1, GX 340+0, 
GX 5-1, GX 17+2 (see more on this issue in  section \S 4).
\par
\noindent
(II). With assumptions that $g\Delta \rho\ll1$ and  $|\nabla p|\ll 1$
the set of equations for $\chi$, $\Upsilon$, $\zeta$ is similar to Eqs (10-12) 
where the right hand side vector is the zero vector, ${\bf 0}=(0,0,0)$.
In this case the dispersion equation for $\omega$  ( for small $\delta<<1$)
\begin{equation}
\omega^2(\omega^2-4\Omega^2)=0
\end{equation}
has the only one nontrivial root $\omega=2\Omega$ which is related to the eigenmode oscillations taking place
parallel to the disk  ($\chi\Upsilon$)-plane.
We discuss the application of this solution to the observation in section \S 4.
\par
\noindent
{\it Case B}
\par
If we do not neglect the pressure gradient and assume the vertical gravitational force and the vector $\bf \Omega$
directed along the vertical ($\delta=0^{0}$), then 
the entire problem (Eqs. 5-9) is reduced to set of equations which have already been  analyzed by Chandrasekhar in C61.
He studied  two cases: (1) In the case of two uniform fluids separated by a horizontal boundary
he showed that for two adjacent, hydrostatic, inviscid fluids,  
the low fluid having density $\rho_1$ and the upper layer having density $\rho_2$ 
the eigenfrequency $\nu_{\rm R-T}$ was 
\begin{equation}
\nu_{\rm R-T}=[2(\Omega/2\pi)^2+(4(\Omega/2\pi)^4+\nu_0^4)^{1/2}]^{1/2}
\end{equation}
where $\nu_0$ is the frequency in the absence of rotation. The frequency $\nu_0$ is of order of 
$\nu_{\rm K}=(g/R)^{1/2}$ if the k-wave number ($\nu_0$ depends on k and
density difference $\Delta\rho=\rho_2-\rho_1$) of order $R^{-1}$ and $\Delta \rho/ (\rho_1+\rho_2)\sim (0.5-1).$ 

For a stratified medium of density $\rho=C_0\exp(\beta z)$, where $1/\beta$  is the scale height, the R-T instability occurs for
positive $\beta$, whereas stable gravity waves occur for negative $\beta$.
The R-T frequency $\nu_{\rm R-T}=\nu_h=[\nu_0^2+4(\Omega/2\pi)^2]^{1/2}$  
if one assumes that the  wave number $k$ is of order of  $d^{-1}\sim H^{-1}$ (where $d$ is a layer size).
It is worthwhile to emphasize that even in the simplest case of the vertical gravitational force (C61)
the hybrid frequency $\nu_h$  is an eigenfrequency (which is also true  for the case with 
no pressure gradient effects, see above).
  Formula (10) for $\nu_{\rm R-T}$  can also be reduced   to $\nu_h$
if one assumes that $\nu_0\sim 2(\Omega/2\pi)$. 
\par
\noindent
{\it Case C}
\par
Now we consider the case when the vector ${\bf\Omega}$ rotates uniformly around the vertical at angle $\delta$, 
then the gravitational vector ${\bf{\cal G}}= \delta \rho(gr/R, 0, zg/R)$, also taking into account the
pressure effects. The general case of ${\bf{\cal G}}= \delta \rho(g_x, g_y, g_z)$ can be analyzed in a similar way 
and it will be presented elsewhere.
 Following the usual practice in problems of this kind, we seek solutions 
of equations (5-9) which are of form (see e.g. C61)
\begin{equation}
u,~v,~w,~\delta\rho.~\delta p={\rm constant}\times\exp(ik_xx+ik_yy+ik_zz+nt),
\end{equation}
where $k_x$, $k_y$ and $k_z$ are the horizontal and vertical wave numbers of the harmonic perturbations respectively.
We also assume that $\rho_0=f(x)\varphi(z)$ is a function of $x$ (or $r$) and $z$ which the scale heights are $1/\gamma$ and $1/\beta$ 
for radial and vertical density profiles respectively, $r\approx R$.
Upon substituting for $u,v,w,\delta\rho,\delta p$ in the form  (16)   equations (5) to (8) become
\begin{eqnarray}
nu-2\Omega_zv=-ik_x(\delta p/\rho_0)+gu\gamma/n+gw\beta/n,~~~~~~~~~~~~~~~~~~\\
nv-2\Omega_xw +2\Omega_zu=-ik_y(\delta p/\rho_0),~~~~~~~~~~~~~~~~~~~~~~~~~~~~~~~~~~\\
nw-2\Omega_xv=-ik_z(\delta p/\rho_0)+(zg/R)u\gamma/n+(zg/R)w\beta/n,~~\\
k_xu+k_yv+k_zw=0.~~~~~~~~~~~~~~~~~~~~~~~~~~~~~~~~~~~~~~~~~~~~~~~~~~~~
\end{eqnarray}
We also use the relation $n\delta \rho+(\gamma u+\beta w)\rho_0=0$ which follows from Eqs. (9) and (16) in order to express 
$\delta \rho$ through $\rho_0$, $u$ and $w$.
The set of equation (17-20)
 assumes a nontrivial solution only if the determinant of the system ${\cal D}=0$. 
This equation provides the dispersion relation for the determination of $n$:
\begin{equation}
P(n)=a_4n^4+a_2n^2+a_1n+a_0=0,
\end{equation}
where $a_0=k_y^2g^2\gamma^2(k_z/k_x-z/R)(\beta/\gamma-k_z/k_x)$,
$a_1=-\beta k_yg(2\Omega_x k_x+2\Omega_zk_z)(1-\gamma z/\beta R)$, 
$a_2= 4(\Omega_xk_x+\Omega_zk_z)^2-g
\beta[-(\gamma/\beta)k_x^2(\beta/\gamma-k_z/k_x)(k_z/k_x-z/R)+k_y^2(\gamma/\beta+z/R)],$~
 $a_4=k^2=k_x^2+k_y^2+k_z^2$.
\subsection{Stable gravity modes}
 
The specific wave values $k_x,k_y,k_z$ are determined by the  conditions imposed on the oscillatory domain boundary.
Thus for a given set of boundary conditions the analysis of the R-T instability is reduced to the analysis of the roots of
of  fourth order algebraic equation (21), which depends on the main parameters of the atmosphere, $\beta$, $\gamma$ and 
ratio of the wave numbers $k_z/k_x$.  We assume that $z/R<1$  and $k_x, ~k_y\propto1/R,~k_z\propto1/z$ is the case of interest.

Figure 1 illustrates the specific behavior of the polynomial P(n) which should help one to
understand the presence (or absence) of its roots for given coefficient $a_0$, $a_1$, $a_2$ 
($a_4>0$, $a_3=0$). For example,
if  $a_0<0$, $a_1<0$ and $a_2> 0$ (see case $i$ below)
then $P(n)$ has two  complex  conjugate roots $n_{1,2}=\zeta\pm i\eta$ and
two real roots $n_3$, $n_4$ (in fact, $d^2P/dn^2>0$). 
Because $a_1<0$, the absolute
value of the positive root $n_4$ is larger than that of the negative one $n_3$. But $a_3$ is a sum
of  the polynomial (real and   complex conjugate) roots,  and  
because of $a_3=n_1+n_2+n_3+n_4=0$  one can come to conclusion that $2\zeta=-(n_3+n_4)<0$. 
Therefore (in this case)  the stable oscillatory mode exists  and $n_{1,2}$ are related to these
damped oscillations. 
 
 Below  we present all cases with the oscillatory stable solutions:
 
 {\it Case i}: $\beta,$ $\beta/\gamma>0$, $\beta/\gamma-k_z/k_x<0$ {\rm and} $a_2>0$ (see Fig. 1, curve i).
For such conditions $a_0<0$ and $a_1<0$. Then equation (21) has two real roots 
which relate to the unstable (growing) and stable (decaying) modes with
one pair of complex  conjugate roots  corresponding to the stable oscillatory mode:
\begin{equation}
n_{1,2}\approx a_1/2d\pm i\omega_h, 
\end{equation}
where $\omega_h=[(|a_2|+d)/2a_4]^{1/2}$ and $d=(a_2^2-4a_4a_0)^{1/2}$. 
These roots of equation (21) are found using  the sequential approximation method: first we solve equation (21)
with $a_1=0$, $\tilde n$ and then in the next stage we look for roots of equation (21) as $n=\tilde n+\alpha$.
  
{\it Case ia}: $\beta,$ $\beta/\gamma>0$, $\beta/\gamma-k_z/k_x<0$ {\rm and} $a_2<0$.
For such conditions $a_0<0$ and $a_1<0$. 
Equation (21) has just one pair of the  complex conjugate roots and  two real roots 
which relate to the unstable (growing) and stable (decaying) modes (see Fig. 1, curve ia).
The oscillatory mode $n_{1,2}$ is stable  for which we have 
\begin{equation}
n_{1,2}\approx a_1/2d\pm i\omega_L, 
\end{equation}
where $\omega_L=[(-|a_2|+d)/2a_4]^{1/2}$. 

{\it Case ib}: $\beta>0,$ $\beta/\gamma<0$, $\beta/\gamma-k_z/k_x<0$. 
For such conditions $a_0<0$ and $a_1<0$ and $a_2>0$.  
This case is similar to case i (see Fig 1, curve i) when  equation (21) has two real roots 
which relate to the unstable (growing) and stable (decaying) modes and 
 complex  conjugate roots which correspond to the stable oscillatory mode:
\begin{equation}
n_{1,2}\approx a_1/2d\pm i\omega_h. 
\end{equation}

{\it Case ii}: $\beta>0$, $\beta/\gamma-k_z/k_x>0$. 
For such conditions $a_0>0$ and $a_1<0$ and $a_2>0$.  
Equation (21) has a pair of  complex conjugate roots (see Fig. 1, curve ii) 
and one of them $n_{1,2}$ is related to  the stable oscillatory mode:
\begin{equation}
n_{1,2}\approx a_1/2d\pm i\omega_h. 
\end{equation}

{\it Case iii}: $\beta<0$, $\beta/\gamma-k_z/k_x>0$, and $a_2, ~d>0$. 
For such conditions $a_0>0$ and $a_1>0$.  
Equation (21) has a pair of complex conjugate  roots (see Fig. 1, curve iii) 
and one of them $n_{3,4}$ is related to  the stable oscillatory mode:
\begin{equation}
n_{1,2}\approx -a_1/2d\pm i\omega_L, 
\end{equation}   
where $\omega_L=[(a_2-d)/2a_4]^{1/2}$.

Finally, we  single out the case when the effective gravitational force goes to zero.
This can happen   during a burst event, when the gravitational forces are compensated by
the radiation pressure forces  (e.g. Titarchuk 1994). 
 For such conditions,  $a_0, a_1 \to 0$ and 
$a_2/a_4=4[(k_x/k)\Omega_x/k+(k_z/k)\Omega_z]^2$.  Equation (21) has  zero roots  $n=0$ and   conjugate complex  roots
\begin{equation}
n_{1.2}=\pm i(a_2/a_4)^{1/2}=\pm 2i[(k_x/k)\Omega_x+(k_z/k)\Omega_z]
\end{equation}
which are related to a pure harmonic mode. In fact, this result also follows from 
treatments detailed in C61 and OT99.
The hybrid frequency then becomes simply $\nu_h=2(\Omega/2\pi)$ for $g=0$.
This case of the perfect coherent oscillations would also occur if the density profile is 
quasi-uniform i.e. when $\beta, ~\gamma\to 0$ but $\beta/\gamma=O(1)$. These two conditions
(on either the effective gravity force or the density profile) can be realized during the burst event. 

In this section the main goal is to reveal all cases when  stable 
gravity modes exist and when the stability breaks down.
This analysis is particularly important in a view of the transient nature of
QPO features (see e.g. Zhang et al. 1998). 
With an increase of bolometric luminosity
(presumably in mass accretion rate) the kHz QPO frequency increases and then 
entirely disappears! At low rates it appears once again!
The stability analysis presented here leads to conditions for the existence and 
the destruction of  gravity modes in terms of density
profiles scale heights $\beta^{-1},~\gamma^{-1}$ and boundary conditions ($k_x,~k_y~k_z$). 
In fact, the strong dependence of the
gravity wave stability on the density profile was a central point of Chandrasekhar's
analysis (C61). For the accretion disk cases,  we can give an
example where  stable  gravity modes are followed by  instabilities (the gravity mode
destruction). 
In case (ib) we have a stable mode with $\omega_h$ as a QPO frequency. When 
the density profile  changes,  over $z-$coordinate from $\gamma<0$ to $\gamma>0$ 
then the stable g-mode with $\omega_h$ can still be sustaned  [see case (ia)],
But if the density profile over z coordinate is stabilized ($\beta<0$) and
that over r coordinate  is inverted (i.e. from $\gamma< 0$ to $\gamma>0$) then the QPO oscillations are
no longer stable. 

\section{Gravity modes and  their relation to QPO phenomenon} 
As we have seen in the previous section, there are two stable oscillatory gravity modes:
one is associated with the hybrid frequency $\omega_h$ (see cases $i,~ib,~ ii$) and another with the low branch frequency 
$\omega_L$  (cases $ia$ and  $iii$). 
We also estimate the decay rate for oscillations as $\lambda=|a_1|/d$ and  the QPO quality  
value $Q=\omega/2\lambda$.
The  presence and absence of these modes  depend on
the atmospheric structure (scale height inverses $\beta$ and $\gamma$) and on the imposed boundary 
conditions (wave numbers $k_x$, $k_y$ and $k_z$).     
Furthermore, it is possible to restore the related boundary conditions if one compares the observed QPO features 
with the calculated mode frequencies and Q-values.  The analysis made in section 3 is also necessary 
in order to control an accuracy of  numerical calculations of the set of hydrodynamical equations (5-9).

To illustrate the results obtained in \S 3  we should specify  the orders of the introduced quantities
$k_x,~k_y,~k_z, \beta, \gamma$.
Namely we assume that  $k_x\sim 1/R,~k_y\sim 1/R,~k_z\sim 1/z,  \gamma\sim q/R$ (where $|q|<1$) 
and $z/R<1.$ If we also assume that $\beta/\gamma-k_z/k_x\sim R/z$ then $\beta\sim 2q/z$.
With these assumptions  expressions for the polynomial coefficients   $a_j$ (see Eq. 21) are significantly simplified:
$a_0/a_4\sim q^2\omega_{\rm K}^4$, $a_1/a_4\sim -2|q|\omega_{\rm K}^2(z\Omega_x/R+\Omega_z)$, $a_2/a_4\sim 
4(z\Omega_x/R+\Omega_z)^2+q\omega_{\rm K}^2$. Because $|q|<1$ and $\omega_{\rm K}<\omega_h$ we can calculate $\omega_L$ as
\begin{equation}
\omega_L\approx |q|(\omega_{\rm K}/\omega_h)\omega_{\rm K}
\end{equation} 
and 
\begin{equation}
Q_L=\omega_L/2(a_1/2d)\approx \omega_h/2\Omega\geq 1/2.
\end{equation}

TOK introduced  the classification scheme for the QPO features 
and  related the observed high and low kHz frequencies  to the hybrid and Keplerian frequencies, $\nu_h=\omega_h/2\pi$ 
and $\nu_{\rm K}=\omega_{\rm K}/2\pi$ respectively. They also attributed the hectohertz frequencies detected 
in the atoll source 4U 1728-34 (Ford \& van der Klis 1999)  to the low branch.  The angle $\delta$ was found to be almost 
twice that found in Sco X-1
(see also T02 for details of $\delta-$ determination). The low frequencies in atoll sources (as a rule)
are three times higher than those in the Z-sources. The hectohertz frequencies have been identified in 
several other neutron star LMXBs (4U 0614+09; van Straaten et al. 2000, 2002; SAX J1808.4-3658, 4U 1705-44;
Wijnands \& van der Klis 1998). They  weakly depend on the kHz frequencies, their ratio to low kHz frequency
being about (4-5) (van Straaten et al. 2002, hereafter S02). The hectohertz frequency Lorentzian profile are very broad
with $Q$-values around 1 or even less. We suggest that these observed frequencies can be identified as
the low branch frequencies (see formulas 28-29). Taking into account  resonance effects (T02), 
we correct the error bars of the hectohertz frequencies  
presented in S02. In Figure 2 the best fit to the data (S02) 
is presented using formula (28) which includes   one fit parameter q. The best-fit value of $|q|=0.3$ 
for which $\chi^2_{red}=0.76$.
The estimated $Q\gax 1/2$ are also  in agreement with the observed values of $Q$ (S02, Table 2).

The harmonic modes $\omega_{hm}= 2(k_x\Omega_x+k_z\Omega_z)$ can be related to a  coherent oscillation frequency of
582 Hz which is observed in 4U 1636-53  during the superburst (Strohmayer \& Markwardt 2002).
If   $k_z/k_x\gg 1$ (which is our case) $\omega_{hm}$ is a double frequency of the NS spin frequency.
Furthermore, we have already found the same  eigenmode with the frequency $2\Omega$ in case (AII) 
when  effects of the buoynant forces and pressure gradients are neglected (see \S 3).
In fact, Miller (1999) has  reported the detection   of the coherent pulsations of frequency  $\sim 291$ Hz 
during  the burst development in 4U 1636-53. {\it Thus  the NS spin frequency $\Omega/2\pi=291 $ Hz 
and the double NS spin frequency $2\Omega/2\pi=582$ Hz have  probably been detected in 4U 1636-53.} 

It is also worth noting that the RXTE observations of  neutron binaries establish
that the kHz QPO frequencies do not seem  to exceed beyond a certain upper limit (Zhang et al. 1998).
This observational effect may be a result of the density profile.  
Our stability analysis presented in \S 3 clearly indicates  such a possibility 
(see also C61). 
\section{Conclusions} 


I have presented a detailed study of the Rayleigh-Taylor (R-T) instability in the accretion flow.
To summarize I have :(1)  put forth  arguments to explain the QPO phenomena, 
as a result of the R-T effect in the rotational frame of reference. (2) formulated and solved the mathematical  problem of
the gravity wave propagation (R-T effect) in the accretion flow. 
(3) concluded that the stable gravity modes in the rotational frame of reference 
are related to the hybrid and low branch frequencies.
(4) demonstrated that  the particular problem of the gravity waves in the rotational frame of reference,
in the approximation of very small pressure gradients, is reduced to the problem of the classical oscillator 
in the rotational frame of reference  which was previously introduced  and applied for the interpretation 
of kHz  QPO observation by OT99. 

I   demonstrate that these frequencies are  intrinsic features of the R-T effect.
They appear in  various configurations of the accretion flow depending on   assumptions 
regarding the density profiles, the boundary conditions and the effects of the pressure forces.
It is not by chance the high and low frequencies phenomenon has common observational appearances for a wide 
range of objects classes from 
black hole sources down to white dwarfs (Mauche 2002). 
(5) Investigated the conditions for the density profile  and the wave numbers (boundary conditions)
 when the gravity modes are stable.
(6) Identified the observed QPO frequencies seen in the power density spectra of NS LMXBs 
using the inferred gravity mode frequencies.
 In particular,  I found that the inferred low branch frequencies and their Q-values are consistent with 
the  QPO hectohertz frequencies observed  in the atoll sources 4U 1728-34 and 4U 0614+19. 
(7) During the NS long (super) burst event, I find that the observer should see oscillations at 
 double NS spin frequency.
The Coriolis force is the only force
which acts in the rotational frame of references and its presence  causes perfect coherent pulsations
with a frequency twice of  the NS spin frequency.

 Finally one can conclude that the R-T gravity wave oscullations must be present and the related QPOs should be
detected in any  system where the gravity, buoynancy and Coriolis force effects
 cannot be excluded (even in the Earth and solar environments).

 L.T. acknowledges  fruitful discussions with Chris Shrader and Kent Wood.
I also appreciate the fruitful discussions with the referee and his/her constructive evaluation  of the manuscript.

\begin{figure} 
\epsscale{1.0} 
\plotone{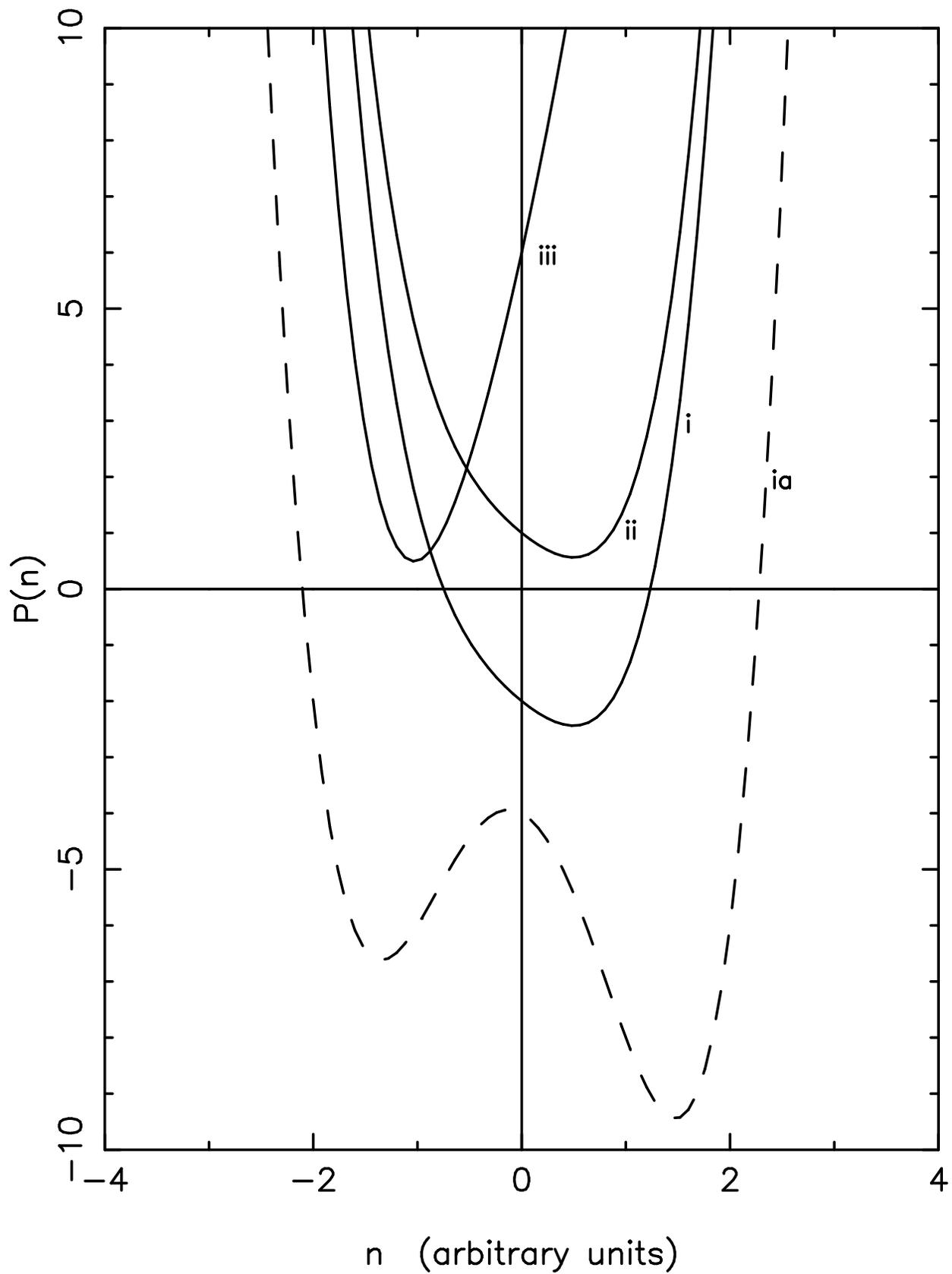} 
\caption{Behavior of the dispersion polynomial P(n) for stable 
gravity modes (see text).}
\label{fig1} 
\end{figure} 

\begin{figure} 
\epsscale{1.0} 
\plotone{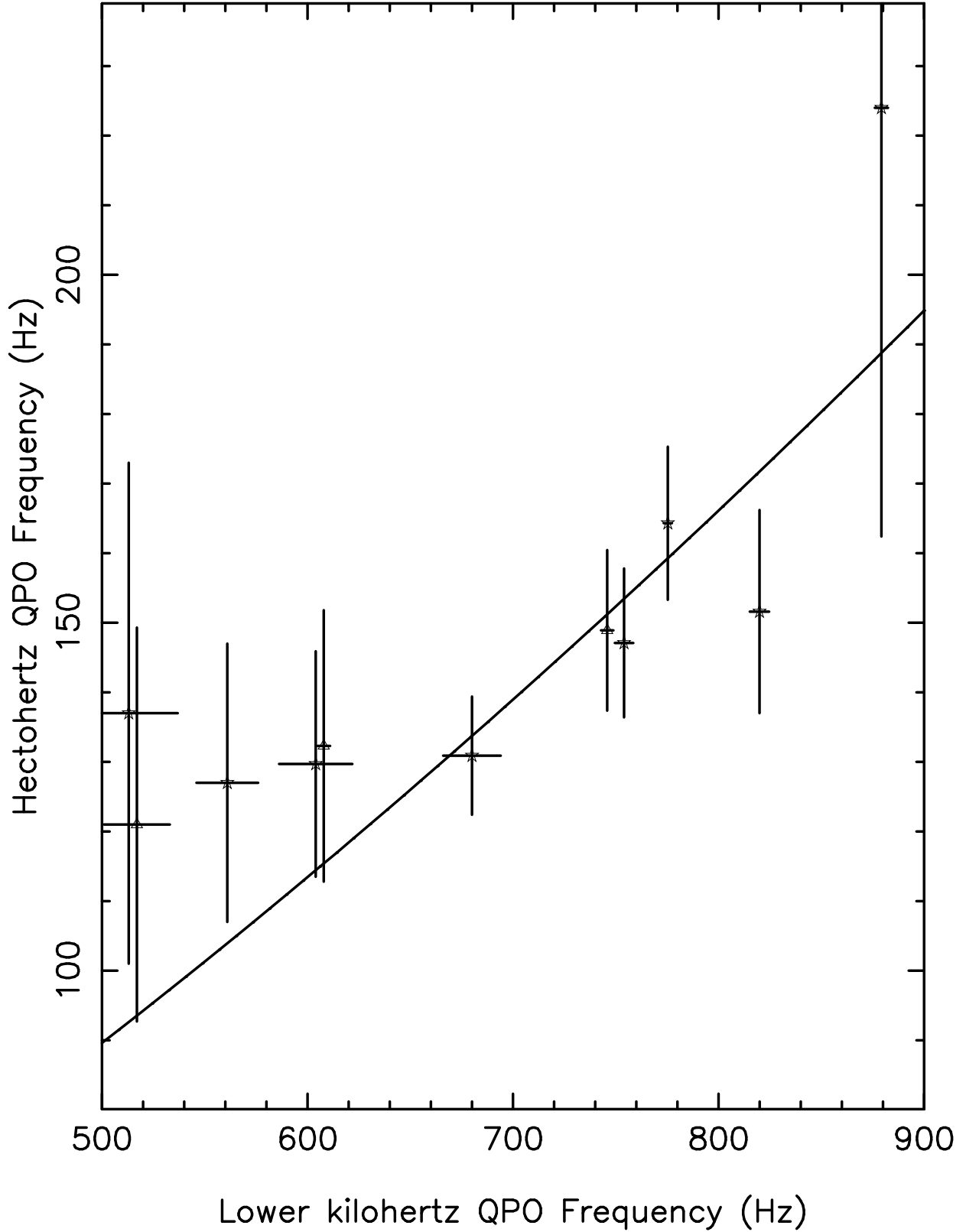} 
\caption{Correlation between the lower kHz frequency and hectohertz frequency
 for 4U 1728-34 (star), 4U 0614+09 (triangles) (van Straaten et al. 2002) and the
best-fit curve of  low branch frequency vs Keplerian frequency 
(solid line).  $\chi^2_{red}=0.76$ for $|q|=0.3$.
  }
\label{fig2} 
\end{figure}

\clearpage

\end{document}